\documentclass[aps,prl,twocolumn,superscriptaddress,showpacs,floatfix]{revtex4}
\usepackage{graphicx}
\usepackage{color}
\begin{document}


\title{Multiple Avalanches Across the Metal-Insulator Transition of Vanadium Oxide Nano-scaled Junctions}


\author{Amos Sharoni}
\email[]{asharoni@physics.ucsd.edu}
\affiliation{Physics Department, University of California-San Diego, La Jolla California 92093-0319, USA}

\author{Juan Gabriel Ram\'irez}
\affiliation{Physics Department, University of California-San Diego, La Jolla California 92093-0319, USA}
\affiliation{Thin Film Group, Universidad del Valle A.A.25360, Cali, Colombia}

\author{Ivan K. Schuller}
\email[]{ischuller@ucsd.edu}
\homepage[]{http://ischuller.ucsd.edu/}
\affiliation{Physics Department, University of California-San Diego, La Jolla California 92093-0319, USA}


\date{\today}

\begin{abstract}
The metal insulator transition of nano-scaled $VO_2$ devices is drastically different from the smooth transport curves generally reported. The temperature driven transition occurs through a series of resistance jumps ranging over 2 decades in amplitude, indicating that the transition is caused by avalanches. We find a power law distribution of the jump amplitudes, demonstrating an inherent property of the $VO_2$ films. We report a surprising relation between jump amplitude and device size. A percolation model captures the general transport behavior, but cannot account for the statistical behavior. 
\end{abstract}


\maketitle


There are many systems in nature that have a transition from one state to another which is driven by an external force, where the transition is not continuous, but rather through a series of avalanches. These systems are diverse as is the driving force and the measurement technique used to observe them. Examples include: Barkhausen noise in ferromagnets \cite{1,2}, acoustic emission in martensitic transition \cite{3,4}, magnetocaloric effect in giant magnetocaloric alloys \cite{5}, sharp magnetization steps or sharp resistance steps in manganites \cite{6,7} and capillary condensation of He in nanoporous material \cite{8}. In general an external parameter modifies the free energy of the two phases, which provides the driving force for the system. The nature of the avalanches provides much information about the system at hand. One can learn about the types of interaction, the role of fluctuations, existence of self organized criticality and the universal features of the transitions which transcend the specific physical system \cite{4,9,10,11}.

Several of these systems are characterized by phase separation during the transition. This implies that the transition occurs through a series of avalanches transforming portions of the system from one phase to the other \cite{12,13,14}. One such system, which has received much attention, is Vanadium Oxide ($VO_2$). $VO_2$ undergoes a first order metal insulator transition (MIT) of over 4 orders of magnitude at $\sim\!340\,K$. The transition is from a high temperature metallic rutile phase to a low temperature insulating monoclinic phase, and can be driven by temperature, light irradiation or pressure \cite{15}. Avalanches may be expected in this system for multiple reasons; it has a first order phase transition, there is a state of phase separation between metallic and insulating regions along the transition \cite{16,17}; and ultra fast measurements reveal a phase transition in separated domains of the system with a transition time on the order of a few picoseconds \cite{18, 19}. In this paper we report the first observation of multiple avalanches across the temperature driven MIT in $VO_2$.

Generally, in order to identify an avalanche, the sensitivity of measurement has to be greater than the magnitude change of the relevant parameter, and the measurement frequency faster than the avalanche frequency \cite{2, 3, 20}. For resistance measurements in $VO_2$ this implies that the size of the device has to be comparable to the magnitude of the domains involved in a single event, and that the rate of change of the driving force slow enough to resolve the avalanches.

By measuring transport properties of $VO_2$ devices, with electrode spacing as small as $200\, nm$ and temperature sweeping rates smaller than $3\, K/min$, we were able to detect that the MIT evolves through a series of discreet jumps ranging over two decades of resistance indicating the transition is through a series of avalanches. We find a scaling law which shows that the amplitudes of the largest jumps are inversely proportional to the device length. Interestingly the jump size distribution follows a power law which is generally robust for various important parameters, including different depositions, device size and temperature sweep rates. The general shape of the transport measurement can be understood in the framework of a non interacting site percolation model. But it cannot explain the statistical nature of the jump sizes, implying that the power law distribution is an expression of an intrinsic property of the $VO_2$ films.

\begin{figure}
\includegraphics[width=8.5cm]{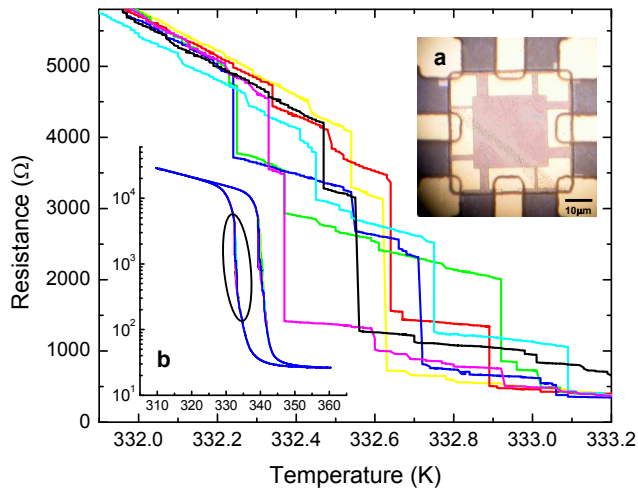}
\caption{\label{fig1} \textbf{Main panel}- 8 consecutive R-T cycles of a $0.5x6\, \mu m^{2}$ $VO_2$ device zoomed in on the area marked in the full measurement shown in \textbf{b}. \textbf{a}- Image of one sample with device length varying between 1 to $6\,\mu  m$ and device width of $6\,\mu  m$.}
\end{figure}

Vanadium oxide thin films were prepared by reactive RF magnetron sputtering of a vanadium target ($1.5"$ diameter, $> 99.8\%$, ACI Alloys, Inc.) on an r-cut sapphire substrate. The samples were prepared in a high vacuum deposition system with a base pressure of $5 \!\times\! 10^{-8}$ torr. A mixture of ultra high purity (UHP) Argon and UHP oxygen gasses were used for sputtering. The total pressure during deposition was $3\!\times\!10^{-3}$ torr, and the oxygen partial pressure was optimized to $1.5\!\times\!10^{-4}$ torr ($5\%$ of the total pressure).  The substrate temperature during deposition was $500^{\circ} C$ while the RF magnetron power was kept at $300\,W$. These conditions yielded a deposition rate of $0.37\, \AA/s$, and a total thickness of $90\, nm$ was deposited for all the samples reported here. The samples were cooled at a rate of $13 ^{\circ}C/min$ in the same $Ar/O_2$ flow of the deposition. Films were characterized and verified to be single phase $VO_2$ by x-ray diffraction (Rigaku RU-200B diffractometer) using Cu K$\alpha$  radiation and Energy-dispersive X-ray spectroscopy (Phillips XL30 ESEM).  Macroscopic measurements of resistance vs. temperature (R-T) curves reveal an MIT starting at $\sim\!340\, K$ with a change of almost 4 orders of magnitude in the resistance, indicative of high purity thin films \cite{21}. 
The $VO_2$ nano-structures were fabricated by standard e-beam lithography and lift-off techniques. On top of the $VO_2$ film metallic electrodes were deposited with $200\, nm$ to $4$ micron separation and $2\, \mu m$ to $15\,\mu  m$ width. The electrodes were deposited by sputtering of $50\, nm$ Vanadium, which acts as an adhesion layer, and $100\, nm$ of gold to assure that the electrode resistance is lower than that of the metallic $VO_2$. A typical sample with 8 devices is depicted in of Fig.\,\ref{fig1}a. Using standard photolithography, the rest of the $VO_2$ was etched in argon plasma, leaving only a square of $VO_2$ (clearly seen in Fig.\,\ref{fig1}a). In the final step the electrodes were photolithographically connected to macroscopic pads. The R-T characteristics were acquired in a home made test-bed by standard 2 probe or quasi 4 probe measurements (with no significant difference between the two) using a constant current source at a measurement rate of $10\, Hz$. Temperature was controlled by a Lakeshore 332 temperature controller, with sweep rates varying between $0.1\, K/min$ and $3\, K/min$. Each measurement was cycled multiple times (10-100) in order to obtain sufficient data for statistical analysis of the jumps.

Figure \ref{fig1} (main panel) shows 8 consecutive cycles of a typical R-T measurement across a $1\!\times\!6\, \mu m^2$ device focusing on the numerous resistance jumps. The full range measurement is depicted in Fig.\, \ref{fig1}b. The transition temperature is $338\, K$ and the total change in resistance is over three orders of magnitude occurring along $\sim\!8\, K$ with a typical hysteresis of $7\, K$, which is similar to the macroscopic properties of the sample. The jumps range from over $2000\, \Omega$  for the largest jumps to below $10\, \Omega$, limited only by the measurement noise. We find that the jumps occur between two consecutive measurements even for the slowest temperature sweep rates, where the (nominal) temperature change between adjacent data points is smaller than $0.5\, mK$. This indicates that the time scale of each jump is much shorter than our measurement capabilities, which is expected due to the fast nature of the $VO_2$ transition, of a few picoseconds \cite{18, 19}. This implies that the transition between the spatially separated but co-existing metallic and insulating phases in $VO_2$ \cite{16, 17} is not continuous, but rather occurs through a series of avalanches, where each avalanche is expressed as a resistance jump in the measurement.

\begin{figure}
\includegraphics[width=8.5cm]{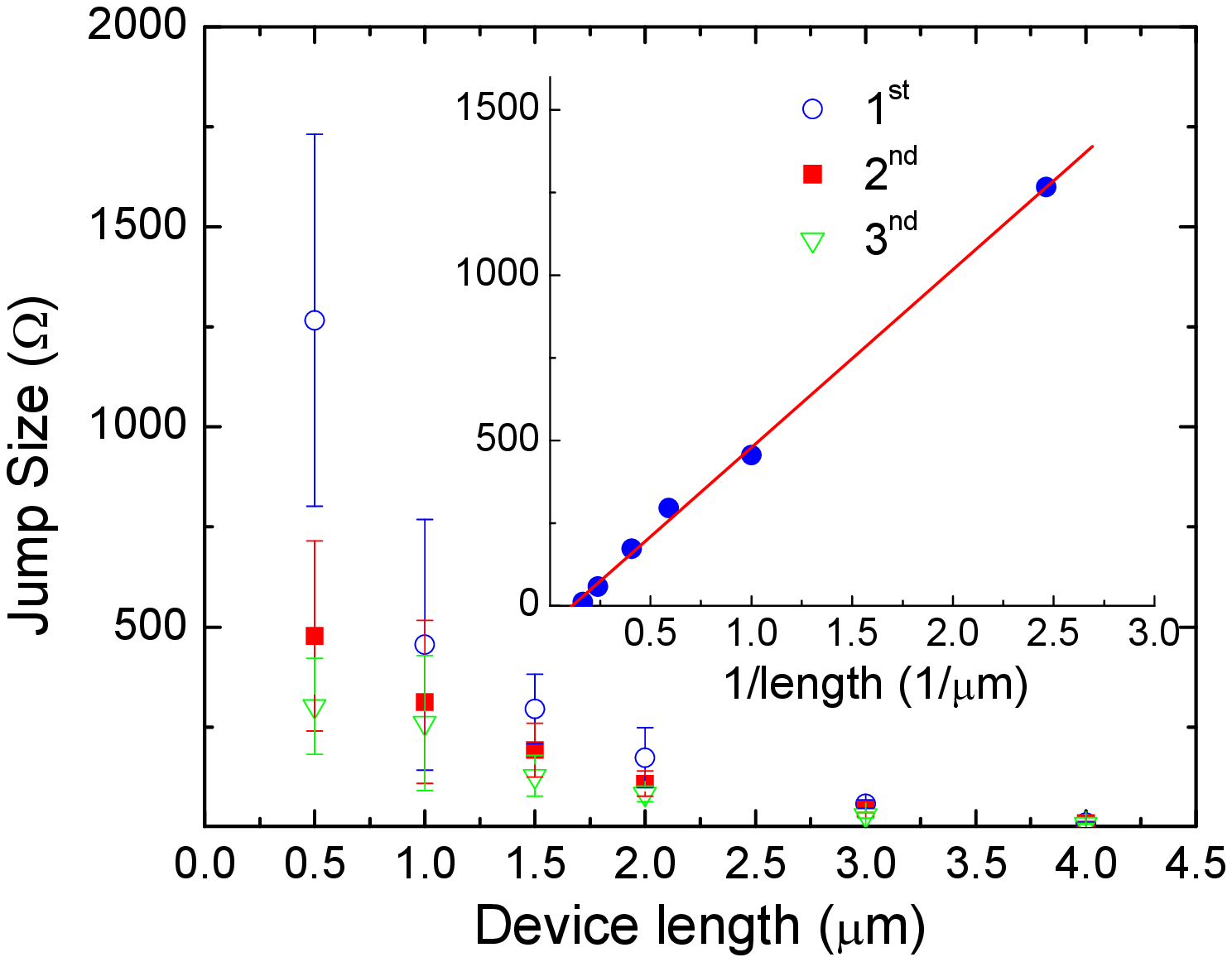}
\caption{\label{fig2} Average value of each of the three largest jumps as a function of electrode spacing. The inset plots only the average of the largest jump as a function of inverse the electrode spacing. Error bars are standard deviations.}
\end{figure}

The main characteristics of the R-T measurement are as follow: the first jumps appear close to the onset temperature of the macroscopic transition. There are 1-3 large jumps (ranging up to a few thousands of ohms in some cases), which may account for $50\%$ of the resistance change. The rest of the jumps are smaller, from a few hundred ohms and are limited by the resolution of our measurement. In some samples, jumps as small as $2\, \Omega$  were resolvable. 
Figure \ref{fig2} shows the three largest jumps (each one averaged over at least 10 cycles and 2 devices) as a function of the device length. Interestingly, the largest jump decreases with device length and is a linear function of inverse the device length, as demonstrated in the inset of Fig.\,\ref{fig2}. We note that the data can be fitted also to an exponential dependence between the largest jump and the device length, but not as well. By extrapolating the linear relation presented in the inset of Fig.\,\ref{fig2} to a jump size of zero we find that the R-T would look smooth for electrode spacing of over 6 microns. This may explain why avalanches were not observed in other experiments performed on devices longer than $10\, \mu m$.

\begin{figure}
\includegraphics[width=8.5cm]{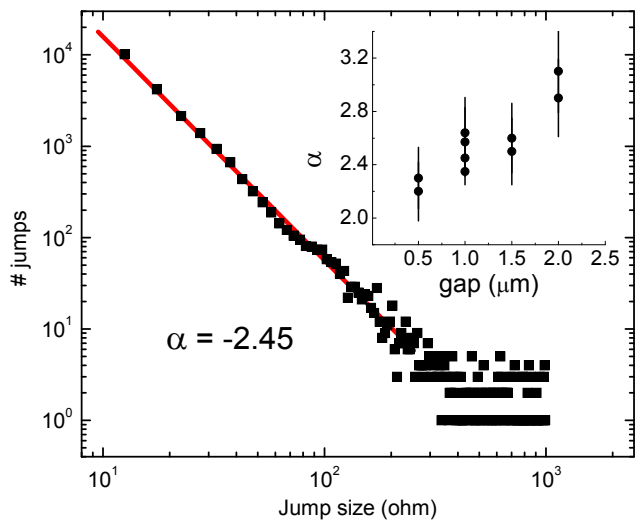}
\caption{\label{fig3} Histogram of the jump amplitudes plotted on a log-log scale, acquired from over 100 cycles measured for a sample $1\!\times\!6\,\mu m^{2}$. Solid line (red online) is a linear fit, indicating power law dependence, with exponent value $\alpha=2.45$. Inset- value of $\alpha$ for different channel lengths and different samples.}
\end{figure}

\begin{figure}
\includegraphics[width=8.2cm]{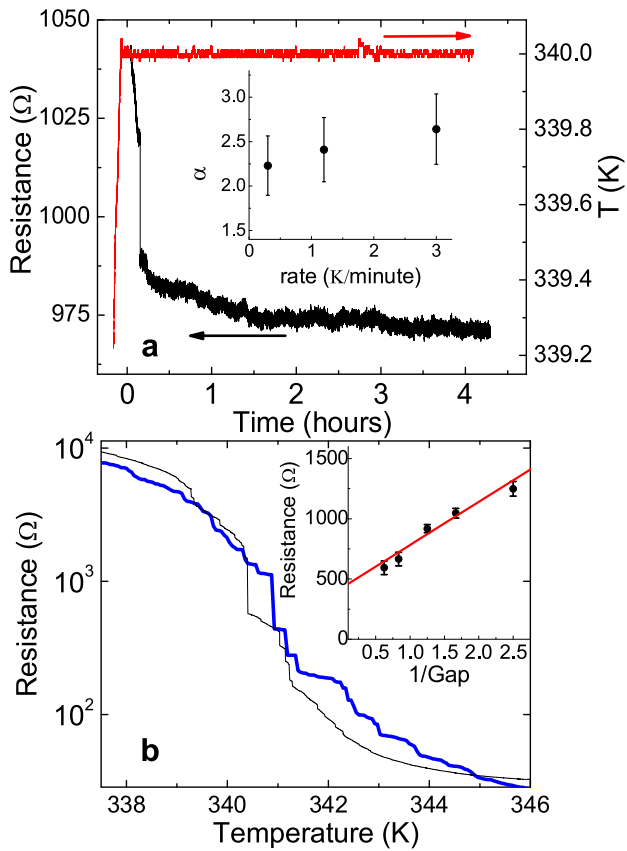}
\caption{\label{fig4} \textbf{a}-	Resistance (bottom curve, left axis) and temperature (top curve, right axis) as a function of time. No avalanches are observed for constant temperatures. Inset- $\alpha$ for different temperature sweep rates. \textbf{b}- Simulation of R-T (thick line, blue online) along with experimental data (thin line, black) from a $1\!\times\!6\, \mu m^{2}$ device. The simulation is of size $20\!\times\!50$ (length $\times$ width) on a square lattice and $\rho_{I}/\rho_{M} = 1000$.}
\end{figure}

The interesting nature of the avalanches is illustrated by the size distribution of the jumps measured along the phase transition. Figure \ref{fig3} shows a histogram of the amplitudes of the jumps, taken from over 100 cycles, measured for a $1\,\mu m$ device at a rate of $1.2\, K/min$. The histogram fits well to a power law, $p(A)\sim A^{-\alpha}$, where $p$ is a probability function and $A$ is the Amplitude of the jump. The value of $\alpha$ from the linear fit is found to be $2.45 \pm 0.1$. The inset of Fig.\,\ref{fig3} depicts exponent values for different electrode spacing. Data for the $1\,\mu m$ device is provided from 3 different samples deposited separately, but under the same nominal conditions. For these the value of $\alpha$  changes between $2.35$ and $2.65$. In addition we find that the power law behavior is identical for heating and cooling, and is also not dependant on the temperature ramp rate, presented in the inset of Fig.\,\ref{fig4}a. To a first approximation, the value of the exponent is very robust in all these samples, and is not strongly affected by any of the parameters tested. Thus, this measured exponent is an inherent and characteristic property of the $VO_2$ thin films.

We find evidence for an athermal transition, i.e. thermal fluctuations do not play an important role in the MIT. For one, as pointed out earlier,   does not vary much as a function of the temperature ramp rate \cite{10}. In addition we monitored the resistance at constant temperatures along the transition. In most cases, like the one presented in Fig.\,\ref{fig4}a, there are no jumps when the temperature is constant. This figure shows that while the temperature is ramped at a rate of $0.2\, K/minute$ for the first 5 minutes (top curve and right axis) there are a number of avalanches and large changes to the resistances (bottom line and left axis). But once the temperature is stable, during the next 4 hours of measurement there are no avalanches. There were a few cases where we did observe a single event even though the temperature was stabilized.

The general characteristics we find in our samples can be interpreted in context of the percolation nature of the measurement system. The observed few large jumps and many smaller jumps with a wide range can be understood in the framework of a percolation network subject to two constraints. First, the resistivity ratio between the insulating and metallic phases is large but finite (in our case  $\rho_I / \rho_M \sim 1000$) and second, the area measured is finite, so the total number of domains which transit through an avalanche is not very large (on the order of  $\rho_I / \rho_M$). In order to model the effects of percolation in the experimental system we preformed numerical simulations of site percolation on a square lattice. This is reasonable, since the macroscopic $VO_2$ MIT can be modeled by an effective medium approximation \cite{22}. In our simulation each site changes randomly from a finite metallic resistance to a finite insulating one and the resistances of all the sites are identical. Figure \ref{fig4}b portrays a simulation with conditions similar to the experimental values. The thick (blue online) line is the simulation, while the thinner one is an experimental R-T. It is evident that this simple toy model captures the main features of the experimental data. We can identify with certainty the largest jump in the experiment as the percolation threshold. If we look at the amplitude of the percolation jump for different device geometries, averaged over 200 realizations, we find that the percolation jump does scale with inverse the device length in a similar fashion to the experiment (shown in the inset of Fig.\,\ref{fig4}b). But, we could not reconstruct the effect of the jump size approaching zero for longer devices.

In the attempt to understand the statistics of the avalanches, it is important to note that the amplitude of a resistance jump cannot be seen as a direct measure of the $VO_2$ volume involved in an avalanche event. Rather, this magnitude is superimposed on the percolation nature of the measurement. This is evident from the wide distribution of jumps seen in the simulations (Fig.\,\ref{fig4}b) even though we assumed that all the sites have identical properties. 
We investigated the contribution of only percolation to the power law distribution of avalanches using the model mentioned above. In order to do that we attempted to calculate values of   for device geometries similar to the experimental one and in the same fashion as we did for the experimental data. Only for part of the data range we could fit the numerical simulations to a power law, since as the jumps become smaller the number of jumps becomes constant and the slope levels. In this range, the value of $\alpha$ varied considerably, between $1.5$ and $3.3$, with simulation size and for different histogram binning choices. This means that the distribution function of the numerical data is not a power law, and the experimental data cannot be explained solely by the geometrical nature of percolation. In order to account for the distribution of avalanches one has to take under consideration intricate conditions of the $VO_2$ system, interactions, inhomogeneities in the sample or in the driving force. This work is beyond the scope of the current paper.


In summary, by choosing a properly nanostructured film we were able to observe discrete transitions in the Metal Insulator transition of $V0_2$ films. These results imply that the metal insulator transition occurs through a series of avalanches. The largest of these discontinuous transitions has an inverse dependence with sample size.  Interestingly, the distribution of resistance jump amplitudes follows a reasonably robust (i.e independent of experimental parameters) power law with an exponent close to $2.4$. This value is similar to the ones reported for acoustic emission measurements of avalanches in martensitic transitions \cite{4, 10}, however in our case the measured amplitude is also a consequence of the percolating nature of the system. Importantly, the power law dependence cannot be explained only by a non-interacting percolation model, meaning that the statistics capture a characteristic property of Vanadium Oxide thin films, which may be a signature of self organized criticality \cite{23}.

\begin{acknowledgments}
We thank Harry Suhl, F\`elix Casanova and Mar\'ia Elena G\'omez for fruitful discussion and Yonatan Dubi for his help with the numerical simulations. This work was supported and funded by the US Department of Energy and the AFOSR.
\end{acknowledgments}

\end{document}